\documentclass[11pt,a4paper]{article}

\usepackage{epsfig,amsmath,amssymb,cite}
\usepackage{hyperref}
\usepackage{graphicx}

\begin{document}

\title{Thin-shell wormholes in (2+1)-dimensional $F(R)$ theories} 
\author{Cecilia Bejarano\thanks{e-mail: cbejarano@iafe.uba.ar}, Ernesto F. Eiroa\thanks{e-mail: eiroa@iafe.uba.ar}, Griselda Figueroa-Aguirre\thanks{e-mail: gfigueroa@iafe.uba.ar}\\
{\small  Instituto de Astronom\'{\i}a y F\'{\i}sica del Espacio (IAFE, CONICET-UBA),}\\
{\small Casilla de Correo 67, Sucursal 28, 1428, Buenos Aires, Argentina}} 
\date{}
\maketitle

\begin{abstract}
We construct a broad family of thin-shell wormholes with circular symmetry in (2+1)-dimensional $F(R)$ theories of gravity, with constant scalar curvature $R$. We study the stability of the static configurations under perturbations preserving the symmetry. We present examples of charged thin-shell wormholes which are asymptotically anti-de Sitter at both sides of the throat. We show that stable solutions are possible when suitable values of the parameters are taken.
\end{abstract}

\section{Introduction}\label{intro} 

Traversable Lorentzian wormholes \cite{whs} have been widely studied in the context of General Relativity (GR) and also in alternative models of gravity. They can represent shortcuts in the same  spacetime or connect two different universes. A quite interesting feature is the fact that they are singularity-free gravitational objects; moreover, wormhole throats can replace black hole singularities in solutions within the framework of some modified gravity theories (see \cite{bej} and references therein).
 
It is well established that wormhole spacetimes within GR always require the presence of exotic matter --which violates at least one of the energy conditions-- somewhere. This is not necessarily the case outside of GR, so it is worth examining the possible violation of the energy conditions by  wormholes in different theories of gravity. The so-called thin-shell wormholes have drawn the attention of the gravitational community, since the exotic matter can be reduced to a minimal amount and be highly confined. In addition to this, the junction procedure also allows to analyze the wormhole dynamics by focusing on the thin shell where the two geometries are joined \cite{whrg,whcil}. Mathematically speaking, these wormholes are built by cutting and pasting two geometries at the throat (located at the shell), where the corresponding junction conditions are accomplished; a geodesically complete manifold is obtained after the matching. This formalism is also applied, for instance, to study thin layers of matter \cite{shrg} around vacuum (namely, bubbles) or black holes. 

The most straightforward generalization of the Einstein-Hilbert dynamical action raises by taking an arbitrary function $F(R)$ of the Ricci scalar $R$, which is usually named $F(R)$ gravity \cite{revfr}. This theory, as well as many other alternatives beyond GR, tries to deal with some of the shortcomings of GR such that the physical meaning of the spacetime singularities and the phenomena associated to the dark sector of the matter and the energy content of the Universe. From the very beginning, $F(R)$ gravity has generated a broad interest; hence, there are many publications in this gravitational framework. Here, we just mention some papers related to the study of black holes by considering constant \cite{bhfr} and non-constant \cite{bhrnonconst} scalar curvature, branes \cite{branefr}, and traversable wormholes \cite{whfr}.

It is well known that the study of thin shells follows in GR the Darmois-Israel scheme \cite{daris}; its counterpart in $F(R)$ gravity is more restrictive \cite{dss, js1}. Besides the continuity of the first fundamental form at the matching hypersurface, for any nonlinear function $F(R)$ the conditions also require the continuity of the trace of the extrinsic curvature and the continuity of the scalar curvature there, with the only exception of quadratic $F(R)$, in which the Ricci scalar can be discontinuous \cite{js1}. In this last case, three new contributions --an external scalar pressure/tension, an external energy flux vector, and a double layer energy-momentum distribution-- emerge in addition to the standard energy-momentum tensor \cite{js1,js2}. The theories with the most generic Lagrangian containing terms quadratic in the curvature \cite{rsv,bdes} share the main attributes with quadratic $F(R)$. Several works can be found within $F(R)$ gravity where the thin-shell formalism is applied to bubbles, layers of matter enclosing black holes \cite{eftsfr, efncrfr}, and also in the construction of traversable wormholes \cite{efwhfr, whtsfr}. The peculiar case of pure double layers in the quadratic $F(R)$ model is studied in Ref. \cite{pdlfr}. A related paper on thin shells within $F(R,T)$ gravity has recently appeared \cite{rosa}.

The interest of $(2+1)$ spacetimes dwells in some of the features which could throw light on conceptual issues related to high energy situations \cite{carlip}, present in the context of black hole singularities, quantum gravity, and string theory. Different three-dimensional geometries were investigated over the years in relation to black holes  \cite{btz} and wormholes \cite{wh3d}. Thin shells of matter \cite{ts3d} as well as thin-shell wormholes \cite{whts3d} were also widely examined. Black hole solutions with $(2+1)$ dimensions in $F(R)$ gravity have been found \cite{hendi14a,hendi14b,hendi20,karakasis}. In the recent article \cite{efwhfr3d}, thin shells of matter in $(2+1)$ dimensions within $F(R)$ theory have been analyzed in detail and examples of stable configurations under radial perturbations have been found.  

In this paper, we use the junction conditions in $F(R)$ gravity to construct circular thin-shell wormholes in $(2+1)$ dimensions.  The scalar curvature $R$ is assumed to be constant. We analyze the stability of the static configurations under perturbations that preserve the symmetry. We consider examples of anti-de Sitter spacetimes with a conformally invariant Maxwell field as a source. In Sects. \ref{jc} and \ref{whs}, we introduce the general formalism, while in Sect. \ref{examples} we show the examples. Finally, in Sect. \ref{discu} we discuss the results obtained. We adopt units such that $c=G=1$, with $c$ the speed of light and $G$ the gravitational constant.

\section{Junction formalism in (2+1)-dimensional $F(R)$ gravity}\label{jc}

The junction formalism allows the construction of a manifold $\mathcal M$ as the result of the union of two parts $\mathcal M_1$ and $\mathcal M_2$ through a hypersurface $\Sigma$, which corresponds to a boundary hypersurface when the matter content vanishes or to a thin shell of matter otherwise. We begin with a brief review of this formalism within $F(R)$ theories in order to apply it to a manifold $\mathcal M$ with (2+1) dimensions, so that the hypersurface $\Sigma$ is one-dimensional.  We denote the first fundamental form (also known as induced metric) on $\Sigma$ by $h^{1,2}_{\mu \nu}$  and the second fundamental form (or extrinsic curvature) by $K^{1,2}_{\mu \nu}$, where the superscripts label each part of $\mathcal M$. The jump of any quantity $\Upsilon $ across the hypersurface $\Sigma$ is defined by $[\Upsilon ]\equiv (\Upsilon ^{2}-\Upsilon ^{1})|_\Sigma$. The matching between $\mathcal M_1$ and $\mathcal M_2$ at $\Sigma$, should fulfill the so-called junction conditions. In $F(R)$ gravity \cite{js1}, the continuity of the first fundamental form inherited from both $\mathcal M_{1,2}$ is required as in General Relativity
\begin{equation}
[h_{\mu \nu}]=0 , 
\label{fffjump}
\end{equation}
but when $F'''(R) \neq 0 $ (where the prime means the derivative with respect to $R$) one has also to demand two additional conditions \cite{js1}: the continuity of the trace of the second fundamental form,
\begin{equation}
[K^{\mu}_{\;\; \mu}]=0 , 
\label{traceKjump}
\end{equation}
and the continuity of the scalar curvature
\begin{equation}
\quad [R]=0 .
\label{rjump}
\end{equation}
In this case, the dynamical equations at the joining hypersurface, with  $S_{\mu \nu}$  the energy-momentum tensor at $\Sigma$, read
\begin{equation}
\kappa  S_{\mu \nu}=-F'(R_\Sigma)[K_{\mu \nu}]+ F''(R_\Sigma)[\eta^\gamma \nabla_\gamma R]  h_{\mu \nu}, \;\;\;\; n^{\mu}S_{\mu\nu}=0,
\label{LanczosGen}
\end{equation}
where $\kappa =8\pi $ and $\nabla$ denotes the covariant derivative. When $F'''(R) = 0$, which corresponds to quadratic $F(R) = R -2\Lambda +\alpha R^2$ gravity, the continuity of the scalar curvature, given by Eq. (\ref{rjump}), is no longer required. Now, the field equations  at $\Sigma$ take the form \cite{js1,js2,rsv}
\begin{equation}
\kappa S_{\mu \nu} =-[K_{\mu\nu}]+ 2 \alpha( [n^{\gamma }\nabla_{\gamma}R] h_{\mu\nu}-[RK_{\mu\nu}]),  \qquad  n^{\mu}S_{\mu\nu}=0.
\label{LanczosQuad}
\end{equation}
However, in the quadratic case, the energy-momentum tensor is not the only matter contribution at $\Sigma$, it is also necessary to take into account three other ones (for more details, see \cite{js1,js2,rsv}): an external energy flux vector $\mathcal{T}_\mu$ 
\begin{equation}
\kappa\mathcal{T}_\mu=- 2 \alpha \overline{\nabla}_\mu[R],  \qquad  n^{\mu}\mathcal{T}_\mu=0,
\label{Tmu}
\end{equation}
with $\overline{\nabla }$ the intrinsic covariant derivative on $(\Sigma,h_{\mu\nu})$;  an external scalar pressure or tension $\mathcal{T}$
\begin{equation}
\kappa\mathcal{T}= 2 \alpha [R] K^\gamma{}_\gamma ;
\label{Tg}
\end{equation}
and a two-covariant symmetric tensor distribution  $\mathcal{T}_{\mu \nu}$
\begin{equation}
\kappa \mathcal{T}_{\mu \nu}=\nabla_{\gamma } \left( 2 \alpha [R] h_{\mu \nu } n^{\gamma } \delta ^{\Sigma }\right),
\label{doublay1}
\end{equation}
with $\delta ^{\Sigma }$ denoting the Dirac delta with support on $\Sigma$, having a resemblance with the dipole distributions in classical electrodynamics \cite{js1,js2,rsv}. These extra contributions are required in order to obtain a well defined energy-momentum tensor with null divergence, which is necessary for local conservation. When $R$ has a null jump at $\Sigma$, all these extra contributions vanish and there is an ordinary thin shell if $S_{\mu \nu} \neq 0$. The typical scenario in quadratic $F(R)$ corresponds to the existence of a double layer besides a thin shell at the joining hypersurface.

\section{Wormholes with a circular throat: construction and stability}\label{whs}

We are interested in metrics with circular symmetry that, adopting the time $t_{1,2}$, the radial $r>0$, and the angular $0\le \theta \le 2\pi$ coordinates, have the form
\begin{equation}
ds^2=-A_{1,2} (r) dt_{1,2}^2+A_{1,2} (r)^{-1} dr^2+r^2d\theta^2 ,
\label{metric-sphe}
\end{equation}
where the subscripts label the regions at the two sides of the matching hypersurface $\Sigma$, defined as a circle with radius $a$. We take the outer region $r \geq a$ of each geometry in order to define the manifolds $\mathcal{M}_{1,2}$. The whole manifold $\mathcal{M}=\mathcal{M}_1 \cup \mathcal{M}_2$ represents  a thin-shell wormhole, with the throat at $\Sigma$, where the flare-out condition is satisfied. The radius of the throat is taken large enough to avoid the presence of event horizons --if the original manifolds have them--, in order to obtain a traversable wormhole. In $\mathcal{M}$ we adopt the coordinates $X^{\alpha }_{1,2} = (t_{1,2},r,\theta)$, while on $\Sigma$ we use the coordinates $\xi ^{i}=(\tau ,\theta )$, with $\tau $ the proper time. We let the radius $a(\tau)$ be a function of $\tau $, and we denote its derivative with respect to $\tau$ by an overdot. The proper time should be the same at both sides of $\Sigma$, then $dt_{1,2}/d\tau = \sqrt{A_{1,2}(a) + \dot{a} ^2}/A_{1,2}(a)$, in which the free signs are fixed by requiring that the times $t_{1,2}$ and $\tau$ all run into the future. The first fundamental form at each side of the shell is given by
\begin{equation}
h^{1,2}_{ij}= \left. g^{1,2}_{\mu\nu}\frac{\partial X^{\mu}_{1,2}}{\partial\xi^{i}}\frac{\partial X^{\nu}_{1,2}}{\partial\xi^{j}}\right| _{\Sigma },
\end{equation}
while the second fundamental form is determined by
\begin{equation}
K_{ij}^{1,2 }=-n_{\gamma }^{1,2 }\left. \left( \frac{\partial ^{2}X^{\gamma
}_{1,2} } {\partial \xi ^{i}\partial \xi ^{j}}+\Gamma _{\alpha \beta }^{\gamma }
\frac{ \partial X^{\alpha }_{1,2}}{\partial \xi ^{i}}\frac{\partial X^{\beta }_{1,2}}{
\partial \xi ^{j}}\right) \right| _{\Sigma },
\label{sff}
\end{equation}
where the unit normals ($n^{\gamma }n_{\gamma }=1$), which are chosen to point from $\mathcal{M}_1$ to $\mathcal{M}_2$, read\footnote{Note that the unit normal $n^\mu$ to $\Sigma$ is well defined without a jump, but for computational purposes the expressions at both sides of $\Sigma$ are usually given.}
\begin{equation}
\label{general_normal_fr}
n_{\gamma }^{1,2 }= \pm \left\{ \left. \left| g^{\alpha \beta }_{1,2}\frac{\partial G}{\partial
X^{\alpha }_{1,2}}\frac{\partial G}{\partial X^{\beta }_{1,2}}\right| ^{-1/2}
\frac{\partial G}{\partial X^{\gamma }_{1,2}} \right\} \right| _{\Sigma },
\end{equation}
with $G(r)\equiv r-a$ (null at $\Sigma$) and the upper and the lower signs corresponding to $\mathcal{M}_1$ and $\mathcal{M}_2$, respectively. On the hypersurface $\Sigma $, we prefer to work in the orthonormal basis $\{ e_{\hat{\tau}}=e_{\tau }, e_{\hat{\theta}}=a^{-1}e_{\theta }\} $, which allows an straightforward interpretation of the results. Then, for the general geometries (\ref{metric-sphe}), the first fundamental form is $h^{1,2}_{\hat{\imath}\hat{\jmath}}= \mathrm{diag}(-1,1)$, the unit normals are
\begin{equation}
n_{\gamma }^{1,2}= \pm \left( -\dot{a}, \frac{\sqrt{A_{1,2}(a)+\dot{a}^2}}{A_{1,2}(a)}, 0 \right),
\end{equation}
and the only non-null components of the extrinsic curvature at each side of $\Sigma$ result
\begin{equation}
\begin{split}
K_{\hat{\tau}\hat{\tau}}^{1,2} & = \pm \frac{A '_{1,2}(a)+2\ddot{a}}{2\sqrt{A_{1,2}(a)+\dot{a}^2}}, \\ 
K_{\hat{\theta}\hat{\theta}}^{1,2} & = \mp \frac{1}{a}\sqrt{A_{1,2} (a) +\dot{a}^2},
\end{split}
\end{equation}
from which we calculate their jumps
\begin{equation}
\begin{split}
[K_{\hat{\tau}\hat{\tau}}] & = -\frac{A '_{2}(a)+2\ddot{a}}{2\sqrt{A_{2}(a)+\dot{a}^2}}-\frac{A '_{1}(a)+2\ddot{a}}{2\sqrt{A_{1}(a)+\dot{a}^2}} ,\\
[K_{\hat{\theta}\hat{\theta}}] & =  \frac{1}{a}\sqrt{A_{2} (a) +\dot{a}^2}+\frac{1}{a}\sqrt{A_{1} (a) +\dot{a}^2} .
\end{split}
\end{equation}
The continuity of the trace of the extrinsic curvature at the shell, i.e. Eq.  (\ref{rjump}), gives the following equation
\begin{equation}
\frac{2\ddot{a}+ A_{2}'(a)}{2\sqrt{A_{2}(a)+\dot{a}^2}}+\frac{2\ddot{a}+ A_{1}'(a)}{2\sqrt{A_{1}(a)+\dot{a}^2}} + \frac{1}{a}\left(\sqrt{A_{1}(a)+\dot{a}^2}+\sqrt{A_{2}(a)+\dot{a}^2}\right) = 0,
\label{cond-dyna}
\end{equation}
which, for the static configurations with a shell radius $a_0$, takes the form
\begin{equation}
\frac{A_{2}'(a_0)}{2\sqrt{A_{2}(a_0)}}+\frac{ A_{1}'(a_0)}{2\sqrt{A_{1}(a_0)}} + \frac{1}{a_0}\left(\sqrt{A_{1}(a_0)}+\sqrt{A_{2}(a_0)}\right) = 0.
\label{cond-stat}
\end{equation}
We now have all the ingredients to find the matter content at the matching hypersurface and then to perform the stability analysis of the thin shell. The energy-momentum tensor in the orthonormal basis has the form  $S_{_{\hat{\imath}\hat{\jmath} }}={\rm diag}(\sigma ,p)$, where $\sigma$ is the energy density and $p=p_{\hat{\theta}}$ is the transverse pressure. In what follows, the scalar curvature is constant in each of the regions at the sides of the throat, then $[\eta^\gamma \nabla_\gamma R]=0$, which simplifies Eqs. (\ref{LanczosGen}) and (\ref{LanczosQuad}), while the so-called \cite{js1} brane tension $\lambda = F''(R_\Sigma) [\eta^\gamma \nabla_\gamma R]$ has a null value at $\Sigma$. 

\subsection{General $F(R)$ gravity with $[R]=0$}

For an arbitrary non-quadratic $F(R)$ theory of gravity, there are three junction conditions which must be satisfied, that is Eqs. (\ref{fffjump}), (\ref{traceKjump}), and (\ref{rjump}). Then, for a constant value $R_0$ at both sides of the throat, from Eq. (\ref{LanczosGen}) we obtain the energy density and the pressure on the shell
\begin{align}
\sigma & = \frac{F'(R_0)}{\kappa }\left( \frac{2\ddot{a}+A_{2}'(a)}{2\sqrt{A_{2}(a)+\dot{a}^2}}+\frac{2\ddot{a}+A_{1}'(a)}{2\sqrt{A_{1}(a)+\dot{a}^2}}\right) ,\\
p & =-\frac{F'(R_0)}{a\kappa}\left( \sqrt{A_{2}(a)+\dot{a}^2}+\sqrt{A_{1}(a)+\dot{a}^2}\right) .
\end{align}
By using the expression (\ref{cond-dyna}), we rewrite the energy density as
\begin{equation}
\sigma=  -\frac{F'(R_0)}{a\kappa}  \left( \sqrt{A_{2}(a)+\dot{a}^2}+\sqrt{A_{1}(a)+\dot{a}^2}\right), 
\end{equation}
so that we can relate it with the pressure by $\sigma -p= 0$. Since the stability analysis is done with respect to the static configurations, we show the corresponding expressions of the energy density and the pressure
\begin{align}
\sigma_0 & = \frac{F'(R_0)}{\kappa}\left( \frac{A_{2}'(a_0)}{2\sqrt{A_{2}(a_0)}}+\frac{A_{1}'(a_0)}{2\sqrt{A_{1}(a_0)}}\right),  \label{energy-stat1} \\
p_0 & =-\frac{F'(R_0)}{a_0\kappa}\left( \sqrt{A_{2}(a_0)}+\sqrt{A_{1}(a_0)}\right).
\label{press-stat1}  
\end{align}
Now we have that $\sigma_0 -p_0=0 $. In quadratic $F(R)$, when $[R]=0$ all these equations are also valid with $F'(R_0)=1+2\alpha R_0$, as it can be easily seen from Eq. (\ref{LanczosQuad}); the extra contributions proportional to $[R]$ vanish in this case.

\subsection{Quadratic $F(R)$ gravity with $[R]\neq 0$}

As it is was stated in Sect. \ref{jc}, the particular case of quadratic $F(R)$ gravity is less restricted: the condition about the continuity of the scalar curvature (\ref{rjump}) is no longer necessary, so we can take constant values $R_1 \neq R_2$ at the sides of the throat. Then, we get from the dynamical equations (\ref{LanczosQuad}) for this case 
\begin{align}
\sigma & = \frac{1+2\alpha R_2}{\kappa }\left( \frac{2\ddot{a}+A_{2}'(a)}{2\sqrt{A_{2}(a)+\dot{a}^2}}\right) +  \frac{1+2\alpha R_1}{\kappa }\left( \frac{2\ddot{a}+A_{1}'(a)}{2\sqrt{A_{1}(a)+\dot{a}^2}} \right) , \\
p & =-\frac{1+2\alpha R_2}{\kappa }\left( \frac{\sqrt{A_{2}(a)+\dot{a}^2}}{a}\right) - \frac{1+2\alpha R_1}{\kappa }\left( \frac{\sqrt{A_{1}(a)+\dot{a}^2}}{a}\right) .
\end{align}
The energy-momentum tensor should be completed by the extra contributions. Namely, the external scalar pressure/tension
\begin{equation}
\mathcal{T}= \frac{2\alpha R_2}{\kappa }\left( \frac{2\ddot{a}+A_{2}'(a)}{2\sqrt{A_{2}(a)+\dot{a}^2}} + \frac{\sqrt{A_{2}(a)+\dot{a}^2}}{a}\right) +  \frac{2\alpha R_1}{\kappa }\left( \frac{2\ddot{a}+A_{1}'(a)}{2\sqrt{A_{1}(a)+\dot{a}^2}} + \frac{\sqrt{A_{1}(a)+\dot{a}^2}}{a}\right) ,
\end{equation}
from which we obtain the relation $\sigma -p=\mathcal{T} $; while the external energy flux vector  results $\mathcal{T}_\mu=0$ and the two-covariant symmetric tensor distribution $\mathcal{T}_{\mu\nu}$ is proportional to $(2 \alpha [R]/\kappa) [R]  h_{\mu \nu}$. The static values of the energy density and the pressure read
\begin{align}
\sigma_0 & = \frac{1+2\alpha R_2}{\kappa }\left( \frac{A_{2}'(a_0)}{2\sqrt{A_{2}(a_0)}}\right) +  \frac{1+2\alpha R_1}{\kappa }\left( \frac{A_{1}'(a_0)}{2\sqrt{A_{1}(a_0)}} \right),
\label{energy-stat2} \\
p_0 & =-\frac{1+2\alpha R_2}{\kappa }\left( \frac{\sqrt{A_{2}(a_0)}}{a_0}\right) - \frac{1+2\alpha R_1}{\kappa }\left( \frac{\sqrt{A_{1}(a_0)}}{a_0}\right) ;
\label{press-stat2}
\end{align}
the extra contributions adopt the form
\begin{equation}
\mathcal{T}_0= \frac{2\alpha R_2}{\kappa }\left( \frac{A_{2}'(a_0)}{2\sqrt{A_{2}(a_0)}} + \frac{\sqrt{A_{2}(a_0)}}{a_0}\right) +  \frac{2\alpha R_1}{\kappa }\left( \frac{A_{1}'(a_0)}{2\sqrt{A_{1}(a_0)}} + \frac{\sqrt{A_{1}(a_0)}}{a_0}\right) ,
\label{extpress-stat}
\end{equation}
which satisfies $\sigma_0 -p_0=\mathcal{T}_0 $; while $\mathcal{T}_\mu^{(0)}=0$ and $\mathcal{T}_{\mu\nu}^{(0)} $ is proportional to $(2 \alpha [R]/\kappa) h_{\mu \nu}$.

\subsection{Stability} 

In  order to analyze the stability of the static circular throats, we employ the standard potential analogy method. By using $\ddot{a}= (1/2)d(\dot{a}^2)/da$, we can rewrite Eq. (\ref{cond-dyna}) to obtain the equivalent equation $az'(a)+z(a)=0$ where $z=\sqrt{A_{2}(a)+\dot{a}^2}+\sqrt{A_{1}(a)+\dot{a}^2}$, which represent a differential equation for $\dot{a}^{2}$ in terms of an effective potential $\dot{a}^{2} = -V(a)$ given by
\begin{equation}
V(a)=-\frac{a^2 \left(A_1(a)-A_2(a)\right)^2}{4 a_0^2 \left(\sqrt{A_1\left(a_0\right)}+\sqrt{A_2\left(a_0\right)}\right)^2}-\frac{a_0^2 \left(\sqrt{A_1\left(a_0\right)}+\sqrt{A_2\left(a_0\right)}\right)^2}{4 a^2}+\frac{A_1(a)+A_2(a)}{2}.
\end{equation}
Since $V(a_0)=0$ and $V'(a_0)=0$, the stability of the configurations under radial perturbations is determined by the positive sign of $V''(a_0)$, which reads
\begin{eqnarray}
V''(a_0) &=& -\frac{3 \left(\sqrt{A_1\left(a_0\right)}+\sqrt{A_2\left(a_0\right)}\right)^2}{2 a_0^2}-\frac{\left(\sqrt{A_1\left(a_0\right)}-\sqrt{A_2\left(a_0\right)}\right)^2}{2 a_0^2} \nonumber \\
&& -\frac{\left(A_1'(a_0)-A_2'(a_0)\right)^2}{2 \left(\sqrt{A_1\left(a_0\right)}+\sqrt{A_2\left(a_0\right)}\right)^2}-\frac{2 \left(A_1(a_0)-A_2(a_0)\right) \left(A_1'(a_0)-A_2'(a_0)\right)}{a_0 \left(\sqrt{A_1\left(a_0\right)}+\sqrt{A_2\left(a_0\right)}\right)^2}  \nonumber \\
&& -\frac{\left(A_1(a_0)-A_2(a_0)\right) \left(A_1''(a_0)-A_2''(a_0)\right)}{2 \left(\sqrt{A_1\left(a_0\right)}+\sqrt{A_2\left(a_0\right)}\right)^2}+\frac{A_1''(a_0)+A_2''(a_0)}{2}.
\label{potd2}
\end{eqnarray}
Now we are in position to explore some concrete examples.

\section{Anti-de Sitter thin-shell wormholes with charge}\label{examples}

In $F(R)=R+ f(R)$ gravity\footnote{Here use the nomenclature in which the first term corresponds to General Relativity.} coupled to nonlinear electrodynamics, the $(2+1)$-dimensional action reads
\begin{equation}
I =\frac{1}{16\pi }\int d^{3}x\sqrt{-g}\left(R+ f(R) +\left( - F_{\alpha\beta}F^{\alpha\beta} \right) ^{s}\right) ,
\label{Action}
\end{equation}
with $s$ an arbitrary positive nonlinearity parameter ($s\neq 1/2$) and $F_{\alpha\beta}F^{\alpha\beta}$ the Maxwell invariant, where $F_{\mu \nu }=\partial _{\mu }A_{\nu }-\partial _{\nu }A_{\mu }$ is the electromagnetic tensor field, being $A_{\mu }$ the gauge potential.  The corresponding energy-momentum tensor has the form $T_{\mu\nu}=1/(4\pi)\left[ -sF_{\mu \gamma}F_{\nu}^{\;\;\gamma}(-F_{\alpha\beta}F^{\alpha\beta})^{s-1}-(1/4) g_{\mu\nu}(-F_{\alpha\beta}F^{\alpha\beta})^s\right]$. In the case with $s = 3/4$, the conformally invariant Maxwell field as a source is obtained \cite{hendi14a}, having a traceless $T_{\mu\nu}$. The field equations in the metric formalism are
\begin{equation}
R_{\mu\nu}(1+f'(R)) - \frac{1}{2}g_{\mu\nu}(R+f(R))+
(g_{\mu\nu}\nabla_\gamma \nabla^\gamma -\nabla_\mu \nabla_\nu)f'(R)=8\pi T_{\mu\nu},
\label{field_eqns}
\end{equation}
\begin{equation}
\partial_\mu \left(\sqrt{-g}F^{\mu\nu}(-F_{\alpha\beta}F^{\alpha\beta})^{-1/4}\right)=0.
\end{equation}
By considering a constant scalar curvature $R_0$, these field equations have a solution of the form (\ref{metric-sphe}), with the metric function \cite{hendi14a} 
\begin{equation}
A(r)=-M-\frac{\left( 2\mathcal{Q}^{2}\right)
^{3/4}}{2\left( 1+f'(R_0)\right) r}-\frac{r^{2}R_{0}}{6},
\label{metric_ads}
\end{equation}
being $M$ the mass and $\mathcal{Q}$ the charge. The only non-null independent component of the electromagnetic field is $F_{tr}=\mathcal{Q}/r^{2}$. When $\mathcal{Q}=0$, the well-known vacuum static BTZ geometry \cite{btz} of General Relativity is obtained, which is also a solution in $F(R)$ gravity \cite{hendi14b} with constant scalar curvature. The spacetime is asymptotically anti-de Sitter since it requires that $R_0<0$. The geometry has a curvature singularity at $r=0$ because the Kretschmann scalar diverges there \cite{hendi14a}. The trace of the field equations 
\begin{equation}
R_0 \left(1+f'(R_0)\right)-\frac{3}{2} \left(R_0+f(R_0)\right)=0 ,
\label{trace1}
\end{equation}
allows to define the effective cosmological constant $\Lambda_e$ 
\begin{equation}
R_0=\frac{3f(R_0)}{2f'(R_0)-1} \equiv 6\Lambda_e .
\label{trace2}
\end{equation}
We can also define an effective charge by
\begin{equation}
Z=\frac{\left( 2\mathcal{Q}^{2}\right)^{3/4}}{2\left( 1+f'(R_0)\right)},
\label{defZ}
\end{equation}
which can be positive or negative, depending on the sign of $F'(R_0)= 1+f'(R_0)$. 
The effective Newton constant $G_{\mathrm{eff}} = G/F'(R) = 1/F'(R)$ is positive when $F'(R)>0$, preventing in this case the graviton to be a ghost \cite{revfr} (see also Ref. \cite{bronnikov}). Note that the sign of $Z$ is fixed by the choice of the $F(R)$ theory and the value $R_0$, while the squared charge $\mathcal{Q}^{2}$ can only modify the absolute value of $Z$. As usual, the radii of the horizons are determined by solving $A(r)=0$ and taking the real and positive solutions\footnote{For $Z\neq0$ the analytic expressions for the solutions of the cubic equations are cumbersome, hence it is meaningless to show them here.\label{cumbersome}}. Then, there are three possible cases:
\begin{itemize}
\item[$-$] If $Z\ge 0$, there is only a solution with $r_h$ the radius of the event horizon; in particular, for $Z=0$ (which corresponds to a quadratic equation) one gets $r_h=\sqrt{-6M/R_0}$. 
\item[$-$] If $Z_c<Z<0$ ($Z_c= (-2M/3)\sqrt{-2M/R_0}$) there are two solutions, one corresponds to the radius of the event horizon $r_h$, and the other to the radius of the inner horizon $r_i < r_h$;  when $Z=Z_c$ both horizons merge into one. 
\item[$-$] If $Z < Z_c$, there are no horizons and the singularity at the origin is naked.
\end{itemize}

In the following, we study some examples by taking into account the metric function (\ref{metric_ads}).  We construct static thin-shell wormholes with radius $a_0$ by applying the junction formalism, and we analyze their stability under radial perturbations which preserve the symmetry. We also determine the kind of matter (ordinary or exotic) located at the thin-shell by checking the weak energy condition (WEC)\footnote{In the orthonormal basis, it takes the form $\sigma_0 \geq 0$ and $\sigma_0 + p_0 \geq 0$.}.

\subsection{Wormhole symmetric across the throat}

\begin{figure}[t!]
\centering
\includegraphics[width=0.9\textwidth]{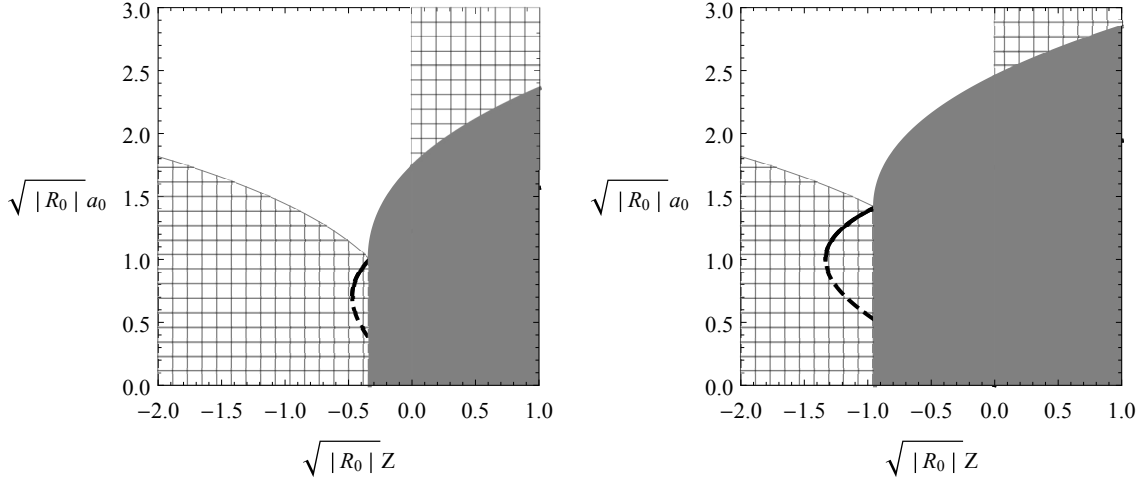}
\caption{Wormhole with the same value of the scalar curvature $R_0$, mass $M$, and charge $\mathcal{Q}$ at the sides of the throat with radius $a_0$, in a general $F(R)$ theory. The parameter $Z=\left( 2\mathcal{Q}^{2}\right)^{3/4}/\left(2\left( 1+f'(R_0)\right)\right) $ has the same sign as $F'(R_0)= 1+f'(R_0)$ (see text). The meshed zones represent normal matter (exotic otherwise) and the gray ones have no physical meaning (see text). Left: $M=0.5$; right: $M=1$. The solid lines correspond to the stable configurations while the dotted lines to the unstable ones.}
\label{wh_R0}
\end{figure}

We proceed with the construction of a thin-shell wormhole symmetric across the throat, that is, with equal values of the scalar curvature $R_1=R_2=R_0$, the mass $M_1=M_2=M$,  and the charge $\mathcal{Q}_1=\mathcal{Q}_2=\mathcal{Q}$, so that the metric functions, given by Eq. (\ref{metric_ads}), are the same $A_1(r)=A_2(r)=A(r)$  in both regions. When $Z \ge Z_c = (-2M/3)\sqrt{-2M/R_0}$, the value of the throat radius $a_0$ is taken larger than $r_h$ in order to remove the region inside the horizon corresponding to the original geometry, while for  $Z < Z_c$ any value $a_0>0$ removes the naked singularity. The proper matching at $\Sigma$ requires that $a_0$ can only take values satisfying Eq. (\ref{cond-stat}). The energy density $\sigma_0$ and the pressure $p_0$ are obtained by replacing the metric functions in Eqs. (\ref{energy-stat1}) and (\ref{press-stat1}), respectively. A configuration is stable under radial perturbations when the second derivative of the potential evaluated at $a_0$, obtained by replacing the metric functions and their derivatives in Eq. (\ref{potd2}), is positive.  

Some representative results are shown in Fig. \ref{wh_R0}. In all plots, the meshed zones represent normal matter satisfying the weak energy condition (exotic matter otherwise) and the gray areas have no physical meaning, corresponding to the removed part of the original manifold. The solid lines correspond to stable static solutions, while the dotted lines to the unstable ones. In the left plot we have adopted $M=0.5$ and in the right one $M=1$. We see that there exist two solutions made of normal matter for a short range of $\sqrt{|R_0|} Z$, corresponding to negative values of $Z$ thus requiring the presence of ghost fields. The solution with the largest radius is stable while the other one is unstable under radial perturbations. A modification in the value of the mass $M$ only results in a change of scale, without affecting the qualitative behavior of the solutions. 

\subsection{Wormhole asymmetric in the mass and the charge}

\begin{figure}[t!]
\centering
\includegraphics[width=0.9\textwidth]{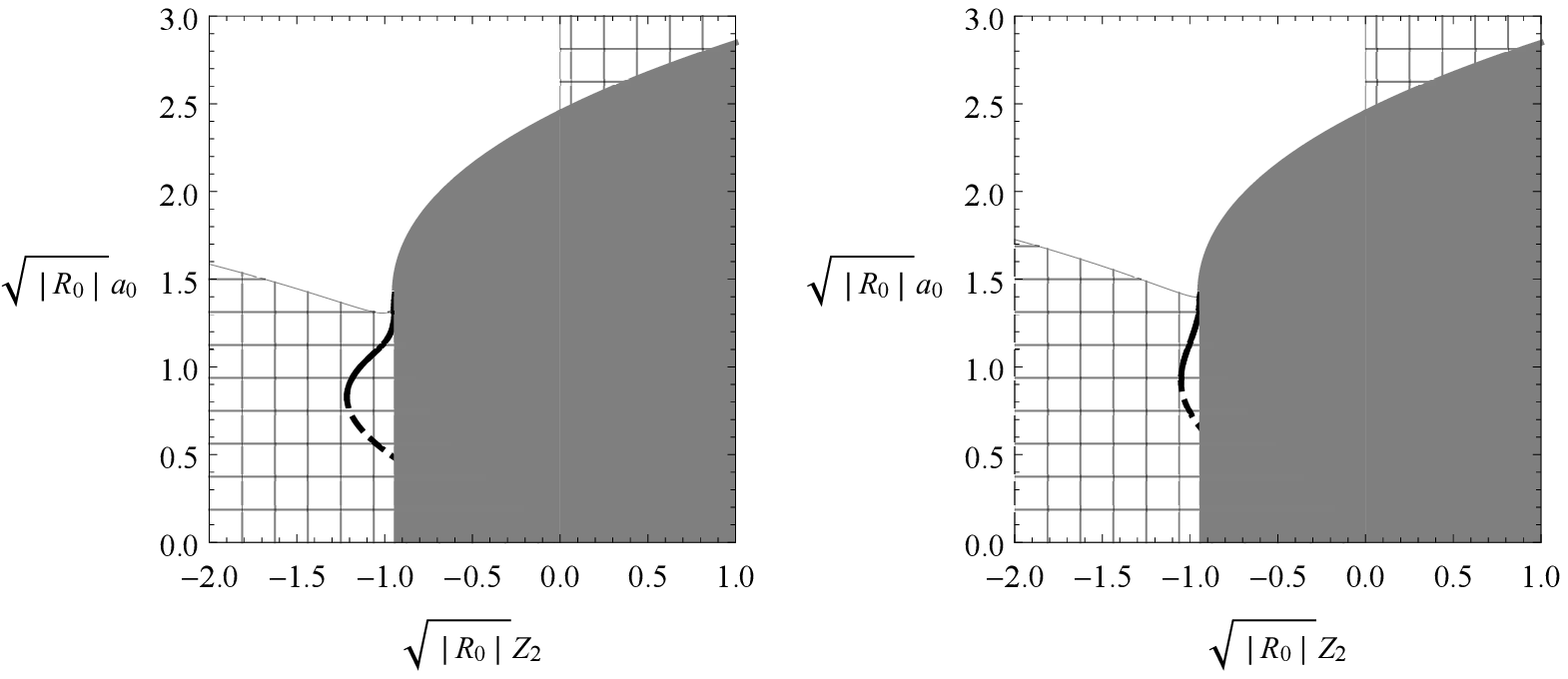}\\
\vspace{0.5cm}
\includegraphics[width=0.9\textwidth]{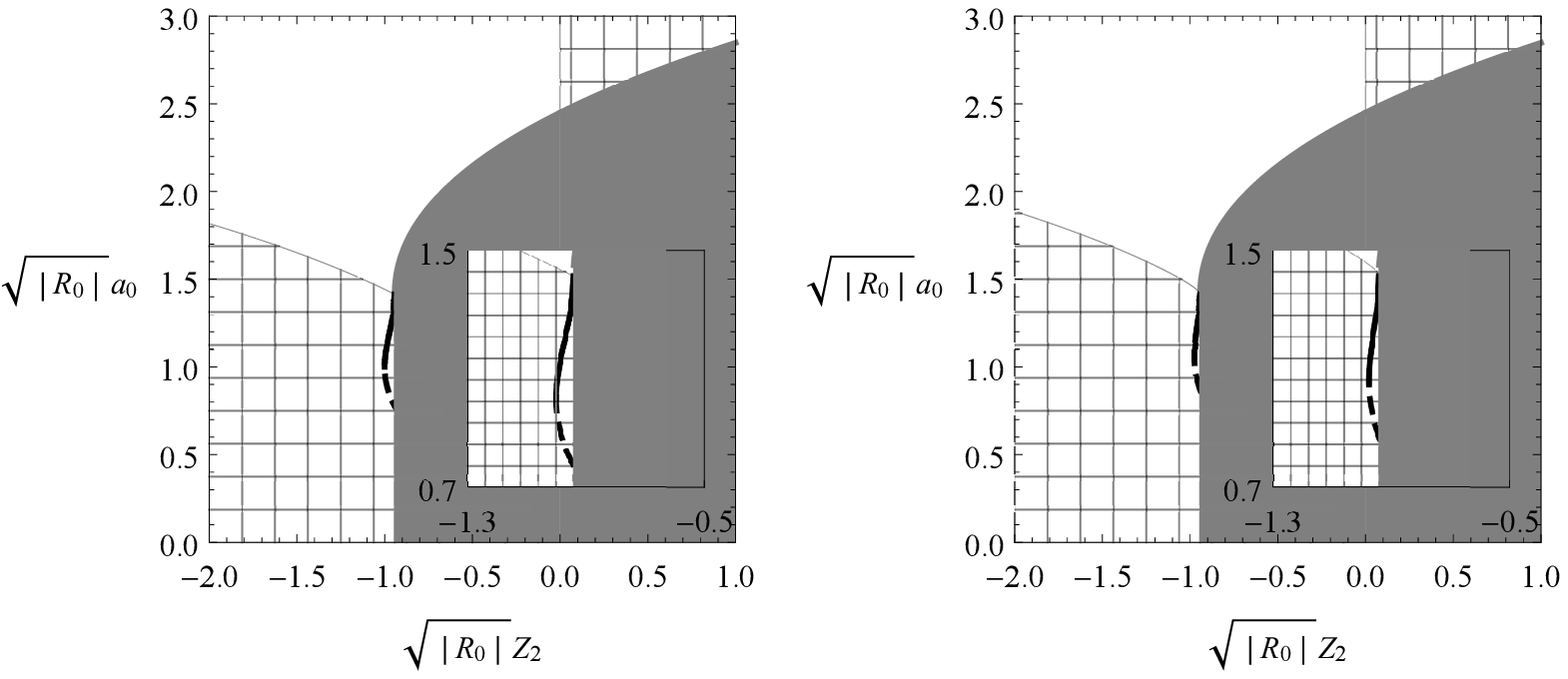}
\caption{Wormhole with the same value $R_0$ of the scalar curvature, but different values of mass $M_1\neq M_2$ at the sides of the throat with radius $a_0$, in any $F(R)$ theory. The parameters $Z_{1,2}=\left( 2\mathcal{Q}_{1,2}^{2}\right)^{3/4}/F'(R_0) $, with $\mathcal{Q}_{1,2}$ the values of charge, have the same sign as $F'(R_0)$ (see text). The solid and dotted lines, and the style of the different zones have the same meaning as in Fig. \ref{wh_R0}. In all plots $M_1=0.5$ and $M_2=1$. Upper row, left: $Z_1 = 0.4 Z_2$;  upper row, right: $Z_1 = 0.7 Z_2$; lower row, left: $Z_1 = Z_2$; lower row, right: $Z_1 = 1.3 Z_2$.}
\label{whM12}
\end{figure}

Now we construct a wormhole with the same value $R_1=R_2=R_0$ of the scalar curvature, but different masses $M_1$ and $M_2$, and charges $\mathcal{Q}_1$ and $\mathcal{Q}_2$ that can be equal or different at the sides of the shell. The metric functions $A_1(r)$ and $A_2(r)$ are given by Eq. (\ref{metric_ads}), in which we use the definitions $Z_{1,2}=\left( 2\mathcal{Q}_{1,2}^{2}\right)^{3/4}/(2\left( 1+f'(R_0)\right)) $. In order to prevent the presence of horizons and singularities, we take the radius of the throat $a_0$ larger than the horizon radii $r_h^{(1)}$ and $r_h^{(2)}$  --when present-- of the original manifolds. The solutions with radius $a_0$ have to satisfy the Eq. (\ref{cond-stat}), and the energy density $\sigma_0$ and the pressure $p_0$ result by replacing the metric functions in Eqs. (\ref{energy-stat1}) and (\ref{press-stat1}), respectively. A configuration is stable under radial perturbations when the second derivative of the potential $V''(a_0)$, shown in Eq. (\ref{potd2}), is positive.

In Fig. \ref{whM12} we show the most illustrative results.  Again, in all plots the meshed zones represent normal matter (exotic otherwise) and the gray areas have no physical meaning, corresponding to the removed parts of the original manifolds. The solid lines correspond to stable static solutions, while the dotted lines to the unstable ones. In all plots the values of the masses are $M_1=0.5$ and $M_2=1$, while the relationship between the charges is $Z_1=\eta Z_2$, with $\eta$ taking the values $0.4$, $0.7$, $1$, and $1.3$, respectively. We can see that in all cases there exist two solutions made of normal matter for a short range of negative values of the charge $Z_2$. This means that solutions are only possible under the presence of ghosts. Regarding their stability, in all cases the solution with the smallest radius is unstable, while the one with the largest one is stable. When the value of $\eta$ increases, the range of $Z_2$ for which the solutions exist becomes smaller. A modification in the values of the non-null masses $M_1$ and $M_2$ causes a change of scale without affecting the qualitative behavior of the solutions.

\subsection{Wormhole asymmetric in the scalar curvature}

\begin{figure}[t!]
\centering
\includegraphics[width=0.9\textwidth]{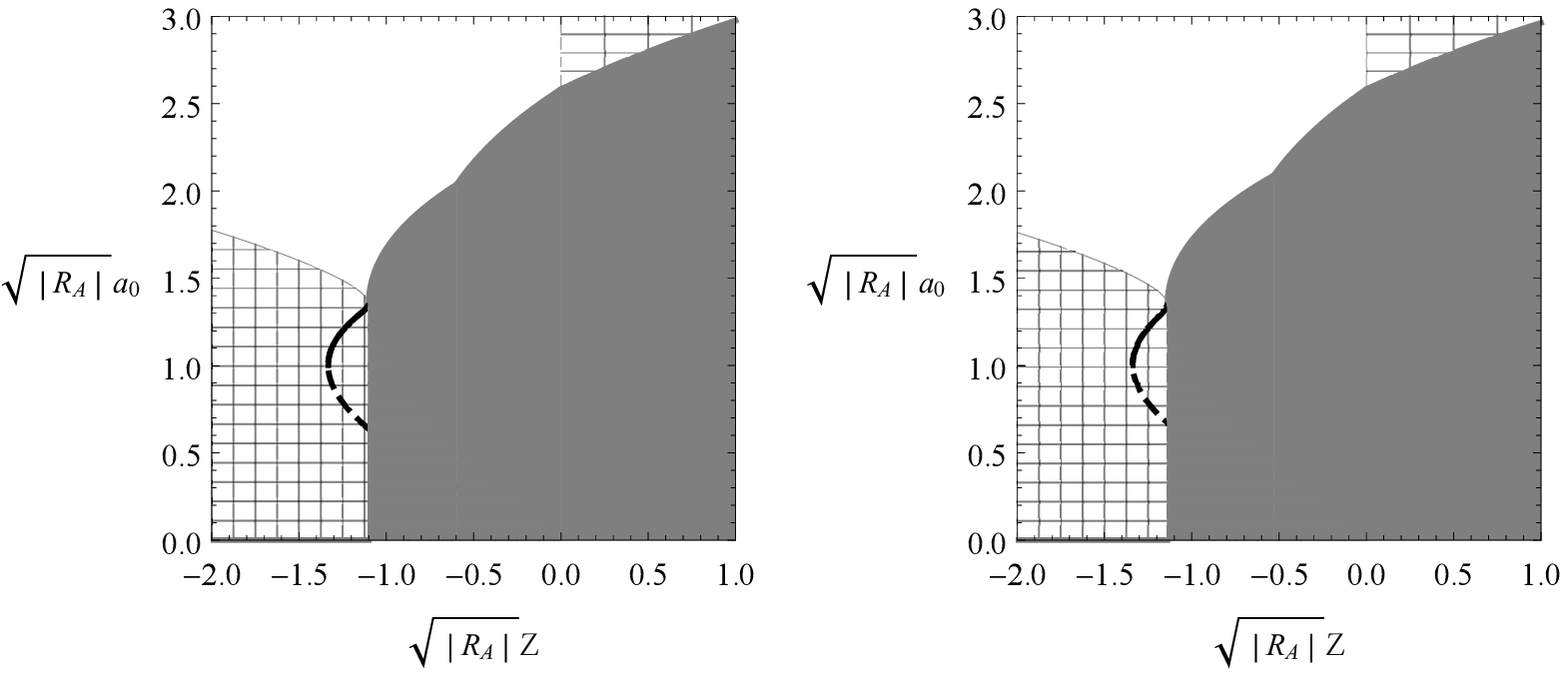}\\
\vspace{0.5cm}
\includegraphics[width=0.9\textwidth]{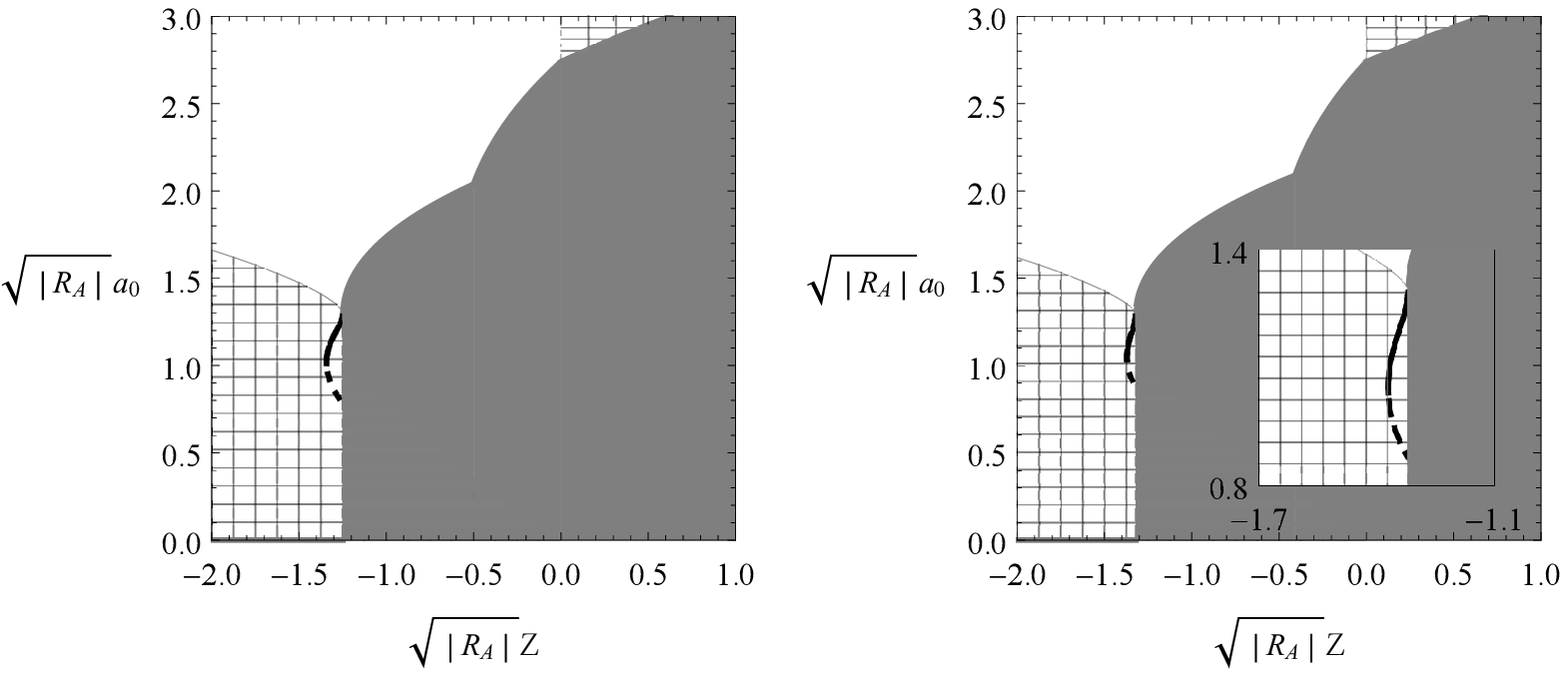}
\caption{Wormhole with different values $R_1\neq R_2$ of the scalar curvature, and the same values of mass $M$ and charge $\mathcal{Q}$, at the sides of the throat with radius $a_0$, only in quadratic $F(R)$ theory. The parameter $Z=\left( 2\mathcal{Q}^{2}\right)^{3/4}/\left(2\left( 1+2\alpha R_A\right)\right) $, with $R_A=(R_1+R_2)/2$, has the same sign as $F'(R_A)= 1+2\alpha R_A$ (see text). The solid and dotted lines, and the style of the different zones have the same meaning as in Fig. \ref{wh_R0}. In all plots $M=1$. Upper row: $R_1 =0.9 R_A$ and $R_2 =1.1 R_A$; lower row: $R_1 =0.8 R_A$ and $R_2 =1.2 R_A$. Left column: $\alpha R_A=-0.1$ for $Z>0$ and $\alpha R_A =-0.9$ for $Z<0$; right column: $\alpha R_A=-0.2$ for $Z>0$ and $\alpha R_A =-0.8$ for $Z<0$}.
\label{wh_R12}
\end{figure}
 
As a final example, we analyze a thin-shell wormhole with equal values of the mass $M_1=M_2=M$ and the charge $\mathcal{Q}_1=\mathcal{Q}_2=\mathcal{Q}$ for both regions, but with different values of the scalar curvature $R_1 \neq R_2$, so that we have to work in quadratic $F(R)$. The metric functions $A_1(r)$ and $A_2(r)$ are then given by Eq. (\ref{metric_ads}), with $R_1$ or $R_2$ as appropriate. For convenience, we define the average value $R_A=(R_1+R_2)/2$, which will be useful in the presentation of the results. We introduce the parameter $Z=\left( 2\mathcal{Q}^{2}\right)^{3/4}/\left(2\left( 1+2\alpha R_A\right)\right) $, which has the same sign as $F'(R_A)= 1+2\alpha R_A$. With this definition, we have that $Z_{1,2}= Z \left( 1+2\alpha R_A\right)/\left( 1+2\alpha R_{1,2}\right)$. As it was previously explained in detail, the throat radius $a_0$ should be larger than the radius of any of the horizons $r_h^{(1)}$ and $r_h^{(2)}$ of the original manifolds in order to remove them and the regions inside, and also satisfy Eq. (\ref{cond-stat}). The energy density $\sigma_0$, the pressure $p_0$, and the external scalar pressure or tension $\mathcal{T}_0$ are calculated by replacing the metric functions in Eqs. (\ref{energy-stat2}), (\ref{press-stat2}), and (\ref{extpress-stat}), respectively; while --as shown above-- there is a null external energy flux vector $\mathcal{T}_\mu$ and the double layer energy-momentum distribution $\mathcal{T}_{\mu \nu }^{(0)}$ is proportional to $(2 \alpha [R]/\kappa) h_{\mu \nu}$. A configuration is stable under radial perturbations when the second derivative of the potential, shown in Eq. (\ref{potd2}), fulfill the inequality $V''(a_0)>0$.  

Some representative results are displayed in Fig. \ref{wh_R12}. The styles of the different regions and the lines have the same meaning explained above. We have taken $M=1$ in all plots, while the values of the scalar curvature are $R_1 =0.9 R_A$ and $R_2 =1.1 R_A$ in the upper row, and $R_1 =0.8 R_A$  $R_2 =1.2 R_A$ in the lower one. In the left column we have adopted $\alpha R_A=-0.1$ for $Z>0$ and $\alpha R_A =-0.9$ for $Z<0$; while in the right one the values are $\alpha R_A=-0.2$ for $Z>0$ and $\alpha R_A =-0.8$ for $Z<0$. In the plots, we find two solutions for a short range of negative values of  $\sqrt{|R_A|} Z$, the one with the largest radius is stable under radial perturbations while the other is unstable; both of them are made of normal matter with the presence of ghost fields. We can also see that the larger the absolute value of the difference between the scalar curvature $R_1$ and $R_2$ across the throat, the smaller the range of values of $\sqrt{|R_A|}Z$ where these solutions exist. A similar behavior can be seen when increasing the difference between the values of the theory parameter $\alpha$ at the sides of the shell, which reduces the range of values of $\sqrt{|R_A|} Z$ where the solutions can be found. A variation in the mass $M$, not illustrated in the figure for brevity, leads again only to a scale change.

\section{Discussion}\label{discu}

In this work, we have presented a wide class of (2+1)-dimensional thin-shell wormholes with circular symmetry, within the framework of $F(R)$ theories of gravity with constant scalar curvature $R$. We have analyzed the matter content at the throat where the thin shell is located and we have studied the stability of the static configurations under perturbations preserving the symmetry.

We have considered three examples of wormholes in $F(R)$ gravity coupled to conformally invariant nonlinear electrodynamics, which are asymptotically anti-de Sitter at both sides of the throat. In the first example, the spacetime is symmetric across the throat, in the second one is asymmetric in the mass and the charge, and in the third one --only in quadratic $F(R)$-- is asymmetric in the scalar curvature. We have obtained that, in all cases, solutions are present only for negative values of $F'(R)$, which means that the presence of ghost fields is always required. In the first example, for symmetric wormholes with fixed (negative) scalar curvature $R_0$, we have found two solutions made of normal matter --satisfying WEC-- for a short range of the squared charge $\mathcal{Q}^2$. The solution with the largest radius is stable while the other is unstable under radial perturbations. Any change in the mass $M$ is reflected as a change of scale conserving the general behavior of the solutions. In the second example, corresponding to the wormhole with the same scalar curvature $R_0$ at both sides of the shell but asymmetric in the mass and the charge, we have also found two solutions with normal matter having the same stability characteristics displayed in the symmetric case. Adopting $M_1 < M_2$, if we increase the ratio $\mathcal{Q}_1^2/\mathcal{Q}_2^2$, the solutions exist for a smaller range of values of charge. Instead, when we modify the masses (maintaining that $M_1 < M_2$) we only see a change of scale without altering the qualitative behavior of the solutions. Finally, in the third example where the wormhole is only asymmetric in the scalar curvature, there are two solutions made of normal matter, with the same stability behavior as in previous examples. When the absolute value of the difference between the constant (negative) scalar curvatures $R_1$ and $R_2$ of the two regions grows, the range of $\mathcal{Q}^2$ where these solutions exist becomes smaller. Analogous behavior can be observed when we decrease the value of the quadratic parameter $\alpha $ of the theory; as it happens with the symmetric case, any change in $M$ results in a change of scale. 

Summarizing, we can say that in our examples of symmetric and asymmetric wormholes across the throat --whether we consider general theories or only quadratic ones--, a pair of solutions, made of normal matter but requiring the presence of ghost fields, exists for a certain set of the parameters of the theory. We have found that under radial perturbations one of these solutions is stable while the other is unstable.

\section*{Acknowledgments}

This work has been supported by CONICET and Universidad de Buenos Aires. C. B. thanks the partial support of the John Templeton Foundation.

\end{document}